\documentclass[twocolumn,trackchanges]{aastex701}
\usepackage{amsmath}
\usepackage{natbib}
\usepackage{graphicx}
\usepackage{booktabs}
\usepackage{float}
\usepackage{CJK}
\usepackage{placeins}
\newcommand{\rev}[1]{#1}
\begin{document}
\begin{CJK*}{UTF8}{gbsn}

\title{Three-dimensional Conditional Diffusion Models for Cosmological 21 cm Lightcone Emulation}

\author[0000-0002-7077-9836]{Bin Xia (夏斌)}
\affiliation{Center for Relativistic Astrophysics, School of Physics,
Georgia Institute of Technology,
Atlanta, GA 30332, USA}
\email[show]{xiabin@gatech.edu}

\author[0000-0003-1173-8847]{John H. Wise}
\affiliation{Center for Relativistic Astrophysics, School of Physics,
Georgia Institute of Technology,
Atlanta, GA 30332, USA}
\email{jwise@physics.gatech.edu}

\begin{abstract}
\rev{We investigate conditional diffusion models for three-dimensional 21 cm
lightcone emulation using $25{,}600$ training lightcones with a sky-plane size
of $64\times64$ and a line-of-sight depth up to 1024 cells. Relative to a
matched 2D baseline, the 3D problem is substantially harder because memory
limits enforce very small micro-batches while the voxel distribution is highly
skewed and long tailed. We compare preprocessing choices, dynamic-range
compression, architecture depth, and training duration. The hyperparameter
sweep is evaluated at one fixed validation point, while the selected
240-epoch 2D and 3D models are evaluated at five validation points using
800 \texttt{21cmFAST} realizations and 800 generated samples per model and
reference set. We assess fidelity with brightness-temperature slices, the
global signal, the power spectrum, and reduced scattering coefficients.
Preprocessing is the dominant factor governing stable training, and
Yeo--Johnson preprocessing with moderate amplitude compression gives the most
favorable trade-off among the tested configurations. The 2D emulator gives
smaller errors in the global signal and reduced scattering coefficients,
whereas the 3D emulator gives smaller central-residual errors in the power
spectrum. Thus, the additional dimension does not yield a uniform improvement
in fidelity under the present architecture and training budget, although it
enables contiguous volumetric generation and improves some two-point
statistics. Measurable biases remain in small-scale and morphology-sensitive
structure, making this work a simulation-level baseline for future
three-dimensional 21 cm emulators with realistic observational effects.}
\end{abstract}
%% Keywords should appear after the \end{abstract} command. 
%% The AAS Journals now uses Unified Astronomy Thesaurus (UAT) concepts:
%% https://astrothesaurus.org
%% You will be asked to selected these concepts during the submission process
%% but this old "keyword" functionality is maintained in case authors want
%% to include these concepts in their preprints.
%%
%% You can use the \uat command to link your UAT concepts back its source.
\keywords{cosmology: theory --- dark ages, reionization, first stars --- methods: numerical --- methods: statistical --- radio lines: general}

%% From the front matter, we move on to the body of the paper.
%% Sections are demarcated by \section and \subsection, respectively.
%% Observe the use of the LaTeX \label
%% command after the \subsection to give a symbolic KEY to the
%% subsection for cross-referencing in a \ref command.
%% You can use LaTeX's \ref and \label commands to keep track of
%% cross-references to sections, equations, tables, and figures.
%% That way, if you change the order of any elements, LaTeX will
%% automatically renumber them.

\section{Introduction}
The redshifted 21 cm signal from the hyperfine transition of neutral hydrogen is a leading probe of Cosmic Dawn and the Epoch of Reionization \citep[EoR;][]{2001PhR...349..125B,2010ARA&A..48..127M,pritchard201221}. In the redshift range $6\lesssim z\lesssim 30$, the signal traces the thermal and ionization state of the intergalactic medium and therefore encodes the impact of the first stars, galaxies, and X-ray sources. Its mean evolution and spatial fluctuations are sensitive to both source physics and large-scale structure, making 21 cm tomography a high-information observable.

A broad observational program is aimed at this regime, including the
Low Frequency Array (LOFAR), the Murchison Widefield Array (MWA), the
Hydrogen Epoch of Reionization Array (HERA), and the future Square Kilometre
Array (SKA) \citep{2013A&A...556A...2V,2013PASA...30....7T,koopmans2015advancing,2017PASP..129d5001D}. Recent HERA upper limits and astrophysical constraints illustrate the broader shift from forecasting toward data-driven inference \rev{\citep{2023ApJ...945..124H, 2026ApJ...998...33A}}. Interpreting future imaging data will therefore require forward models that are both fast and physically reliable across broad astrophysical parameter spaces.

\rev{Semi-numerical simulations such as \texttt{21cmFAST} are widely used to generate target distributions for such analyses 
\citep{Mesinger_2007,2011MNRAS.411..955M,Murray2020}. 
In this work, we use \texttt{py21cmfast} v3.3.1 for the training and validation 
lightcones.} However, large simulation campaigns remain expensive when repeated realizations, parameter sweeps, or downstream simulation-based inference are required. 
This has motivated growing interest in machine-learning emulators for 21 cm cosmology.

Machine learning has already been applied to parameter inference from 
tomographic data, lightcone compression, foreground-affected reconstruction, 
and fast sample generation 
\citep{10.1093/mnras/stab3215,10.1093/mnras/stab1158,2021ApJ...907...44V,Zhao_2022}. 
Emulators based on generative adversarial networks (GANs) have also been 
explored for 21 cm data products 
\citep{2014arXiv1406.2661G,10.1093/mnras/staa523,2022arXiv221105000A,2023arXiv230704976D}, 
but these models remain sensitive to training instability, partial mode collapse, 
and preprocessing choices. 
\rev{The range of emulated targets is also broad, including recurrent
emulators of the one-dimensional global 21 cm signal
\citep{DorigoJones_2024}, generative models for line-intensity-mapping mock
lightcones \citep{10.1093/mnras/staf2124}, and diffusion- or
recurrent-network-based emulators of 21 cm summary observables
\citep{2025A&A...703A.130B,breitman202621cmemuv3}.}
\rev{Field-level emulation of reionization has followed several distinct
strategies. CRADLE uses two-dimensional slices of the gas-density and
stellar-number-density fields to predict maps of the reionization time, which
are subsequently combined to reconstruct a three-dimensional reionization
field \citep{2019MNRAS.490.1055C}. PINION instead predicts the
four-dimensional evolution of the hydrogen ionization fraction from smoothed
gas- and matter-density fields and supplements its data-driven objective with
a constraint derived from the reionization chemistry equation
\citep{2023MNRAS.521..902K}. More recently, CosmoUiT uses a
transformer--U-Net hybrid to predict coeval three-dimensional neutral-fraction
fields from dark-matter and halo-density fields together with reionization
parameters, from which brightness-temperature cubes are constructed
\citep{2026JCAP...06..011P}. These approaches emulate physically informative
field-to-field mappings, but the realization-specific density or source
fields supplied to the network already constrain much of the spatial
structure of the desired output.}

\rev{Generative lightcone studies make a different set of compromises.
\citet{2025ApJ...992..173D} use a conditional StyleGAN2 model and
multi-fidelity transfer learning to generate large-scale two-dimensional
transverse--line-of-sight lightcone images from astrophysical parameters and
random latent variables. In contrast, \citet{2026A&A...706A..15H} synthesize
full three-dimensional lightcones with a maximum-entropy model constrained by
3D scattering coefficients measured from a target lightcone. The latter
approach retains a three-dimensional output but represents the target through
compressed, morphology-sensitive summaries and is not trained as a
parameter-conditioned emulator spanning a broad astrophysical parameter
space. These studies illustrate complementary ways of reducing the difficulty
of field generation, including restricting the output dimensionality,
supplying realization-specific physical fields, or constraining the output
through compressed summary representations.}

\rev{Diffusion models provide a complementary approach to stochastic
high-dimensional generation without adversarial training
\citep{ho2020denoising,dhariwal2021diffusion}. The closest generative
precedent for the present work is \citet{zhao2023can}, who conditionally
generated two-dimensional 21 cm brightness-temperature images from
astrophysical parameters and compared DDPM and StyleGAN2 using
scattering-based distributional diagnostics and Fisher forecasts. Related
studies have further shown that different 21 cm summary statistics retain
complementary information and that wavelet-scattering-based summaries can
outperform power-spectrum-only analyses in some inference settings
\citep{10.1093/mnras/staa3165,2024ApJ...973...41Z,2024A&A...688A.199P}.}

\rev{The present work occupies a different point in this design space. We ask
a diffusion model to generate a full three-dimensional brightness-temperature
lightcone conditioned only on two global astrophysical parameters, together
with stochastic diffusion noise. No realization-specific density, halo,
source, or ionization field is provided, and the target lightcone is not
replaced by a compressed representation. The network must therefore learn
both the parameter-dependent evolution of the signal and the distribution of
spatial structure induced by varying initial conditions. The challenge is
consequently determined not only by the dimensionality and volume of the
output, but also by the limited amount of realization-specific information
supplied at the input. A model may reproduce visually plausible slices while
still retaining measurable biases in the global signal, power spectrum, or
higher-order morphology.}
% In this paper, we address three main questions. \rev{First, can a conditional diffusion model remain 
% stable and useful when extended to full 3D lightcones, and how does its 
% performance compare with a matched 2D baseline?} 
% % \rev{This comparison is important because 3D lightcones contain line-of-sight evolution and spatial coherence that are absent from isolated 2D slices, but
% % they are also substantially harder to train. The additional dimensionality may
% % therefore either provide useful physical structure or simply make the learning problem too difficult under fixed computational resources.}
% Second, which preprocessing and dynamic range control choices are most important for stable training? Third, what levels of physical fidelity are recovered across a hierarchy of diagnostics probing mean evolution, two-point structure, and higher-order non-Gaussian morphology? We will show that the answer is not captured by image inspection alone. Across the tested configurations, the most favorable behavior arises only after explicit control of transformed-space skewness and dynamic range, and the achieved fidelity follows a clear hierarchy: the agreement is strongest in the global signal, weaker in the power spectrum, and weakest in higher-order scattering statistics.

\rev{In this paper, we address three main questions. First, can a conditional
diffusion model remain stable and useful when extended to full 3D lightcones,
and how does its performance compare with a matched 2D baseline? Second, which
preprocessing and dynamic-range-control choices are most important for stable
training? Third, what levels of physical fidelity are recovered across
diagnostics probing mean evolution, two-point structure, and higher-order
non-Gaussian morphology? We find that the matched 2D model is easier to
optimize and performs better for the global signal and reduced scattering
coefficients, while the 3D model gives smaller central-residual errors in the
power spectrum. Thus, the additional dimension provides a
diagnostic-dependent rather than universal advantage. Across both
dimensionalities, the global signal is recovered most accurately, while
measurable discrepancies remain in the power spectrum and scattering
coefficients.}

The remainder of this paper is organized as follows.
Section~\ref{sec:background} summarizes the physical and methodological background, emphasizing the morphology of the 21 cm signal and why three-dimensional conditional diffusion is more challenging than the two-dimensional setting.
Section~\ref{sec:methods} describes the simulation data, conditioning parameters, validation design, diffusion architecture, preprocessing pipeline, and evaluation protocol.
% Section~\ref{sec:results} presents the hyperparameter sweep, identifies the preferred preprocessing and dynamic range control choices, and examines the representative 3D model across image-space and summary-statistic diagnostics.
% Section~\ref{sec:discussion} discusses what the experiments demonstrate, what limitations remain, and what is required before such emulators can be used in survey-facing inference pipelines.
% Section~\ref{sec:conclusion} summarizes the main conclusions and outlines the role of this work as a simulation-level baseline for future 3D 21 cm emulator studies.
\rev{Section~\ref{sec:results} compares the matched best 2D and 3D models
across image-space, global-signal, power-spectrum, and scattering-coefficient
diagnostics. Section~\ref{sec:discussion} discusses the resulting
dimensionality--fidelity trade-off, the remaining limitations, and the steps
required before such emulators can be used in survey-facing inference
pipelines. Section~\ref{sec:conclusion} summarizes the main conclusions.
Appendix~\ref{app:hparam} presents the full preprocessing and hyperparameter
study, including the voxel-distribution comparisons used to select and
diagnose the preferred models.}

\section{Physical and Methodological Background}
\label{sec:background}

\subsection{21 cm Morphology and Non-Gaussian Structure}
The redshifted 21 cm signal provides a tomographic view of the thermal and ionization history of the intergalactic medium prior to cosmic reionization. Its morphology is shaped by several coupled physical processes, including density evolution, radiative coupling, gas heating, and reionization. After thermal decoupling from the cosmic microwave background (CMB), the gas cools adiabatically and can produce a 21 cm absorption signal. Once the first luminous sources appear, Ly$\alpha$ coupling drives the spin temperature toward the gas temperature, strengthening the absorption feature, while subsequent X-ray heating can move the signal into emission. During reionization, ionized bubbles expand and suppress the 21 cm brightness temperature over progressively larger regions until most hydrogen becomes ionized \citep{2001PhR...349..125B,pritchard201221}.

These processes imprint structure on multiple physical scales and generate strongly non-Gaussian spatial patterns. In some regions of parameter space, the 21 cm field remains close to a Gaussian-like density field, whereas in others it develops prominent bubble morphology, sharp ionization boundaries, and strongly nonlinear contrasts. This combination of large-scale evolution and non-Gaussian structure is precisely what makes 21 cm tomography scientifically powerful, but it also makes emulation demanding: visually plausible samples must reproduce not only mean trends but also morphology-sensitive summaries \citep{zhao2023can}.

\subsection{From 2D to 3D: Motivation and Cost}

\rev{The closest diffusion-model precedent for the present work is the 2D
21 cm slice emulator of \citet{zhao2023can}, which considered conditional
generation of $64\times64$ brightness-temperature images at individual redshift
slices rather than contiguous 3D lightcones. 
Their results demonstrated that
conditional diffusion can reproduce 21 cm image statistics in a lower-dimensional
slice setting. \rev{Relative to these isolated fixed-redshift images, both
lightcone representations used in our matched comparison retain the
line-of-sight evolution. Relative to our transverse--line-of-sight 2D
baseline, the 3D representation specifically adds the second transverse
dimension and the correlations among neighboring transverse--line-of-sight
planes. This extension enables contiguous volumetric realizations with
spatial coherence across both transverse directions while retaining the
non-stationary progression of heating and reionization along the line of
sight.}
At the same time, this extension is not automatically
beneficial: the additional dimension greatly increases the output size, memory
cost, training I/O, and optimization difficulty. 
We therefore treat the move from a transverse--line-of-sight 2D
representation to a full 3D lightcone as an empirical question: the additional
dimension contains physically meaningful transverse coherence, but it also
makes the conditional distribution substantially more difficult to learn
under a fixed computational budget.
This comparison must also distinguish output dimensionality from
conditioning information: predicting a 3D field from realization-specific
density and source fields is a more tightly constrained mapping than
generating the corresponding distribution from global astrophysical
parameters alone.
}
In this higher-dimensional setting, transformed-space behavior and
dynamic-range control become much more consequential because memory limits force
very small micro-batches and amplify optimization noise. The present problem is
harder for three coupled reasons.

First, the computational burden grows sharply. Full-resolution 3D lightcone training must contend with strong GPU-memory limitations, small micro-batches, and large training I/O. Second, those computational limits interact with the statistics of the data: heavy-tailed voxel distributions, strong skewness, and inverse-transform distortions become harder to control when optimization noise is amplified by small effective batch sizes. Third, evaluation becomes more demanding. A model can reproduce the broad visual progression of slices while still failing to match lower-order or higher-order summary statistics. These coupled effects motivate the normalization-centered and diagnostic-centered study developed below.

\section{Methods}
\label{sec:methods}

\subsection{Data, Conditioning, and Validation Design}
Our target field is the differential 21 cm brightness temperature relative to the CMB,
\begin{equation}
T_{21}(\mathbf{x}, z)=\tilde{T}_{21}(z)\,x_{\rm HI}(\mathbf{x})\,[1+\delta(\mathbf{x})]\left(1-\frac{T_{\gamma}}{T_S}\right),
\end{equation}
where
\(\tilde{T}_{21}(z)
=
27
(\Omega_b h^2/0.023)
\sqrt{(0.15/\Omega_m h^2)(1+z)/10}\)
is the redshift-dependent brightness-temperature normalization in units of mK.
$x_{\rm HI}$ is the neutral hydrogen fraction, $\delta$ is the matter overdensity, $T_{\gamma}$ is the CMB temperature, and $T_S$ is the hydrogen spin temperature \citep{furlanetto2006cosmology}. 
\rev{Because the lightcones contain
both absorption and emission, we retain the spatial and redshift evolution
of the spin temperature rather than adopting the high-spin-temperature
approximation $T_S\gg T_{\gamma}$. The resulting brightness-temperature
field therefore reflects the coupled evolution of the density, neutral
fraction, and spin-temperature fields.}

Our training dataset has a shape of $(25600,64,64,1024)$ for the 25600 3D lightcones and $(25600,2)$ for the conditioning parameters. The corresponding lightcones span 2048~cMpc (comoving megaparsecs) along the line of sight, with a redshift range $z\in[7.51,20.95]$ and comoving distance range $[8.973,11.021]$~cGpc. Each voxel therefore corresponds to 2~cMpc along the line of sight, while the transverse dimensions each span 128~cMpc. The conditioning variables are \((\log_{10}T_{\rm vir}, \zeta)\),
where \(T_{\rm vir}\) is the minimum virial temperature of halos capable
of hosting ionizing sources and \(\zeta\) is the ionizing efficiency,
which controls the number of ionizing photons produced per collapsed baryon.
In our training set, \(\log_{10}T_{\rm vir} \in [4,6]\) and
\(\zeta \in [10,250]\). Following \citet{zhao2023can}, these two parameters are chosen because they capture dominant astrophysical variation in the 21 cm signal while keeping the conditional problem tractable.

The training parameter points are drawn with Latin hypercube sampling across the target parameter space, with one simulated lightcone for each parameter pair.
\rev{Each training lightcone is generated with a distinct, independently assigned 
\texttt{py21cmfast} random seed, so the initial conditions vary across the 
training database rather than being held fixed.}
This design is chosen to maximize parameter space coverage at fixed simulation cost. The cosmological parameters are fixed to $(\Omega_m,\Omega_b,n_s,\sigma_8,h)=(0.310,0.0490,0.967,0.810,0.677)$ with the variables having their standard definitions \citep{ade2016planck}.

For validation, we instead fix several parameter points and generate many realizations at each point. 
\rev{As in the training set, different validation realizations are generated 
with distinct \texttt{py21cmfast} random seeds, so the validation ensembles sample initial-condition variance at fixed astrophysical parameters.}
Specifically, we generated five reference parameter sets, each with lightcone shape $(64,64,1024)$, at
$(\log_{10}T_{\rm vir},\zeta)=$ $(4.4,131.341)$, $(5.477,200)$, $(4.699,30)$, $(5.6,19.037)$, and $(4.8,131.341)$.
These points were chosen to sample physically distinct regimes within the
training domain. Lower values of \(T_{\rm vir}\), associated with lower impact of feedback, allow lower-mass halos to
host ionizing sources, increasing the abundance of contributing sources,
whereas larger \(\zeta\) corresponds to more efficient ionizing photon
production per collapsed baryon. Thus, the low \(T_{\rm vir}\),
high \(\zeta\) point \((4.4,131.341)\) represents a regime with relatively
abundant and efficient ionizing sources, while the high \(T_{\rm vir}\),
low \(\zeta\) point \((5.6,19.037)\) represents the opposite limit.
The points \((4.699,30)\) and \((5.477,200)\) correspond approximately to
the ``Faint Galaxy'' and ``Bright Galaxy'' benchmark models of
\citet{Greig_Mesinger_2017}, respectively, and the intermediate point
\((4.8,131.341)\) provides an additional visually structured validation case within the parameter space.
For each parameter set, we produced 800 \texttt{21cmFAST} realizations.
\rev{We did not perform a dedicated convergence study to determine the
minimum ensemble size required for each statistic. The choice of 800 realizations follows
the validation setup of the 2D diffusion-emulator study of
\citet{zhao2023can} and provides a conservative reference ensemble with
reduced Monte Carlo noise in the reported medians, percentile bands, and
residual metrics.}
This multi-realization validation design is intended to test not only whether 
the diffusion model reproduces the central tendency of the signal at fixed 
parameters, but also whether it learns the conditional diversity of the target 
distribution.

\rev{For the hyperparameter sweep that is detailed in Appendix \ref{app:hparam}, each run is evaluated at the fixed
validation point
$(\log_{10}T_{\rm vir},\zeta)=(4.699,30)$ using 800 diffusion samples,
which are compared with the corresponding 800 \texttt{21cmFAST}
realizations. The matched 2D--3D models corresponding to indices 5, 6, 14,
and 15 in Table~\ref{tab:job_registry} in Appendix \ref{app:hparam} additionally sampled at all five
validation points. The main comparison of the best 240-epoch models in
Figures~\ref{fig:single_job_global_power}
and~\ref{fig:single_job_scattering} therefore uses all five parameter points,
whereas the hyperparameter metrics in Table~\ref{tab:job_registry} and
Figure~\ref{fig:hparam_mae} use the single fixed validation point.}

\rev{All five validation points lie within the training range, so the present
validation deliberately tests interpolation within the training domain rather
than extrapolation beyond it. We did not evaluate extrapolative performance
in this work; doing so would require additional validation simulations outside
the training parameter range and is left for future study.}

Our code supports 2D and 3D input tensors through the same interface:
\begin{equation}
\mathbf{x}\in \mathbb{R}^{H\times Z} \;\text{(2D)}\quad \text{or}\quad
\mathbf{x}\in \mathbb{R}^{H\times H\times Z} \;\text{(3D)}.
\end{equation}
In this work, $H=64$ and $Z$ varies by run design, with the most demanding configurations reaching $Z=1024$.
\rev{The 2D representation retains one transverse direction together with
the full line-of-sight evolution, whereas the 3D representation retains both
transverse directions. We construct two matched 2D--3D model pairs:
indices 5 and 15 in Table~\ref{tab:job_registry}, trained for 120 epochs,
and indices 6 and 14, trained for 240 epochs. Within each pair, the models
use the same Yeo--Johnson preprocessing, amplitude factor $A=0.1$, and one
residual block per encoder level, and are matched in training duration while
differing in dimensionality. The 120-epoch pair provides a controlled
comparison within the hyperparameter sweep, while the 240-epoch pair is used
for the main comparison in
Figures~\ref{fig:single_job_global_power}
and~\ref{fig:single_job_scattering}.}

\subsection{Conditional Diffusion Architecture}
We use a context-conditioned U-Net \citep{ho2020denoising, dhariwal2021diffusion} in which timestep embeddings and condition embeddings are combined in latent space. For the $64\times64\times1024$ 3D lightcone datasets, the network uses four resolution levels with an anisotropic downsampling and upsampling stride of $(2,2,4)$, producing feature resolutions of $(64,64,1024)$, $(32,32,256)$, $(16,16,64)$, and $(8,8,16)$. The fiducial comparison architecture uses 128 base channels and channel multipliers $(1,1,2,4)$. We apply one residual block per resolution in the encoder and two residual blocks per resolution in the decoder; additional tests with deeper residual-block configurations are included in the hyperparameter sweep.

The model is dimension agnostic through shared 2D/3D code paths built from \texttt{Conv\{2,3\}d} in PyTorch, residual blocks, downsampling and upsampling operators, and attention blocks. Attention is applied with four heads at the two lowest spatial resolutions, corresponding to feature maps of $(16,16,64)$ and $(8,8,16)$ in the 3D case. The timestep embedding dimension is set to 512, corresponding to four times the base channel size. The two astrophysical conditioning parameters are mapped through a linear token embedding of the same dimension, and the condition embedding is added directly to the timestep embedding before it is injected into every residual block.

% For all reported experiments, the sampling follows the full denoising diffusion probabilistic model (DDPM) trajectory with 1000 diffusion timesteps and a standard mean-squared-error (MSE) noise-prediction loss \citep{ho2020denoising}. 
For all reported experiments, training uses a standard mean-squared-error (MSE) noise-prediction objective with 1000 diffusion timesteps, while sampling follows the full denoising diffusion probabilistic model (DDPM) trajectory \citep{ho2020denoising}.
We use the AdamW optimizer with an initial learning rate of $10^{-5}$ and a cosine-annealing learning-rate decay. 
% For the fiducial 3D runs, each GPU processes 800 lightcones, corresponding to the full training set of 25,600 lightcones distributed across 32 GPUs. 
The training set of 25,600 lightcones is distributed evenly across 32 GPUs, so each GPU is assigned 800 lightcones per epoch.
\rev{For the 120-epoch 3D runs, we use a micro-batch size of 2 and gradient
accumulation of 16, corresponding to a cosine schedule with
$T_{\max}=3000$ optimizer steps. For the matched 240-epoch 2D and 3D models
used in the main comparison, the cosine schedule is extended proportionally
to $T_{\max}=6000$ optimizer steps.}
We also tested an exponential moving average (EMA) with a decay rate $\beta=0.995$ but found no improvement in the reported comparisons; all results shown in this paper therefore exclude EMA.

\subsection{Normalization and Dynamic-Range Control}
\label{subsec:normalization}
% A central empirical finding of this study is that preprocessing dominates training stability at 3D scale. 
\rev{Preprocessing is especially important in the 3D lightcone problem
because the raw brightness-temperature distribution is highly asymmetric,
strongly concentrated near \(T_b\simeq 0\), and long tailed toward large
negative absorption values. The consequences of these distributional features
are amplified by the large output volume: small transform-space biases can
affect many voxels and become visibly important after inversion back to physical
units. In addition, memory limits force very small micro-batches, making the
optimization more sensitive to poorly conditioned voxel distributions. The
preprocessing choices below should therefore be viewed as problem-specific
feature engineering that imposes an inductive bias matched to the 21 cm
lightcone distribution.}
The full preprocessing and postprocessing pipeline is
\begin{align*}
& x_{\rm raw}\rightarrow \text{primary transform}\rightarrow \text{optional squish}\rightarrow \\
&\text{diffusion model} \rightarrow \text{inverse squish}\rightarrow \text{inverse transform},
\end{align*}
where \(x_{\rm raw}\) denotes the physical 21 cm brightness-temperature voxel value, and the optional squish step refers to the additional dynamic-range compression defined in Eq.~(\ref{eq:squish}).
We consider four primary normalization transforms:
\begin{enumerate}
\item \textbf{Yeo--Johnson transform}:
\begin{align}
x'&=\mathrm{YJ}(x;\lambda) \nonumber \\
&=
\begin{cases}
\dfrac{(x+1)^\lambda-1}{\lambda}, & \lambda\neq 0,\ x\ge 0,\\[6pt]
\log(x+1), & \lambda=0,\ x\ge 0,\\[6pt]
-\dfrac{(1-x)^{\,2-\lambda}-1}{2-\lambda}, & \lambda\neq 2,\ x<0,\\[6pt]
-\log(1-x), & \lambda=2,\ x<0,
\end{cases}
\end{align}
followed by standardization
\begin{align}
\tilde{x}=\frac{x'-\mu_{\mathrm{YJ}}}{\sigma_{\mathrm{YJ}}},
\end{align}
where $\lambda$ is the fitted Yeo--Johnson parameter and $(\mu_{\mathrm{YJ}},\sigma_{\mathrm{YJ}})$ are the mean and standard deviation of the transformed training distribution.

\item \textbf{Min--max scaling}:
\begin{equation}
\tilde{x}=2\,\frac{x-x_{\min}}{x_{\max}-x_{\min}}-1,
\end{equation}
which maps the data to $[-1,1]$ using fixed dataset-level bounds.

\item \textbf{Z-score normalization}:
\begin{equation}
\tilde{x}=\frac{x-\mu}{\sigma},
\end{equation}
with fixed dataset mean $\mu$ and standard deviation $\sigma$.

\item \textbf{Arcsinh transform}:
\begin{equation}
\tilde{x}=\mathrm{arcsinh}(x).
\end{equation}
\end{enumerate}

The physical voxel distribution motivates these choices. Most voxels cluster near a brightness temperature of $0$~mK; however, there is an additional concentration in the emission phase in positive temperatures, and the absorption tail extends more broadly toward large negative values. This strong skewness makes transformed-space behavior critical: a transform can appear reasonable in transformed space while still producing clear distortions after inversion back to physical units.

After the primary normalization transform, we optionally apply an additional dynamic-range compression step (``squish'') before diffusion training:
\begin{equation}
\label{eq:squish}
x_{\mathrm{sq}}=\begin{cases}
A\,\tilde{x}, & k=0,\\
A\,\mathrm{arcsinh}(\tilde{x}/k), & k\neq 0.
\end{cases}
\end{equation}
% In practice, the main 3D comparisons in this paper use the linear case $k=0$, so the squish step reduces to a simple amplitude scaling $x_{\rm sq}=A\tilde{x}$. We tested several amplitude settings and found that $A=0.1$ provides the most favorable trade-off among the configurations considered here, outperforming both weaker compression ($A=0.5$, $A=1$) and overly aggressive compression ($A=0.01$). Tests with nonlinear squish, such as $(A,k)=(0.1,1)$, gave behavior similar to the linear $(0.1,0)$ case. We therefore interpret the main role of the squish stage as controlled dynamic-range compression rather than as a qualitatively new nonlinear transform.
\rev{For the matched 2D--3D comparison presented in the main text, we use the linear
case $k=0$ with $A=0.1$, so that $x_{\rm sq}=0.1\tilde{x}$. The full comparison
of primary transforms, amplitude factors, residual-block depths, and training
durations is presented in Appendix~\ref{app:hparam}. That sweep motivates the
configuration adopted for the main dimensional comparison.}

\begin{figure*}[!htbp]
\centering
\includegraphics[width=\textwidth]{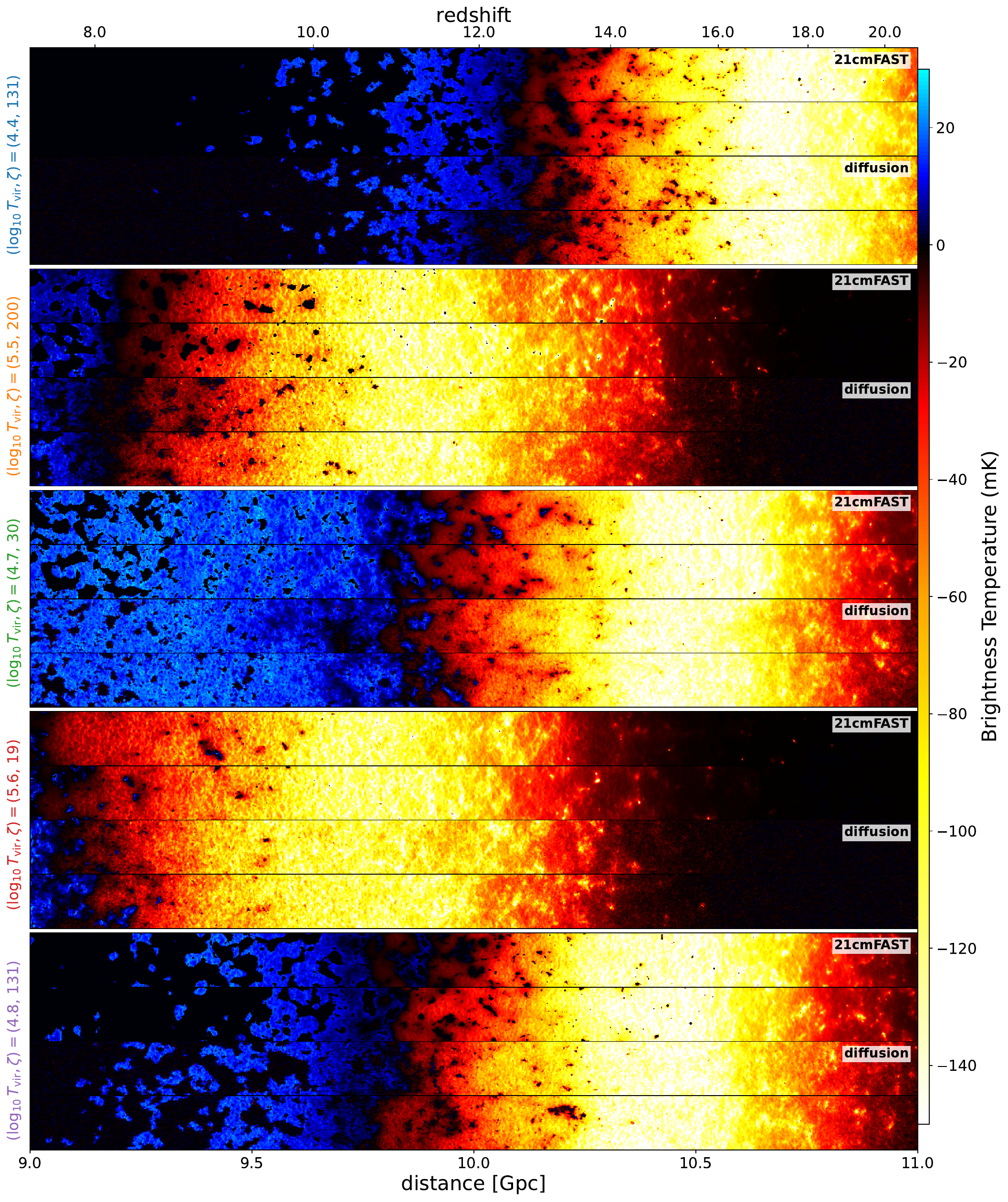}
\caption{
\rev{Brightness-temperature slice comparison for the best 3D model used in
the matched dimensional comparison (Yeo--Johnson preprocessing, $A=0.1$, one
residual block per encoder level, 240 epochs; Table~\ref{tab:job_registry},
index 14). For each conditioning point, the upper two strips show independent
\texttt{21cmFAST} realizations and the lower two strips show
diffusion-generated realizations. Figures~\ref{fig:single_job_global_power}
and~\ref{fig:single_job_scattering} compare this model quantitatively with the
matched best 2D model.}
}
\label{fig:single_job_tb}
\end{figure*}
% \FloatBarrier

\subsection{Evaluation Protocol}
We evaluate generated samples against \texttt{21cmFAST} targets using a hierarchy of complementary diagnostics:
\begin{enumerate}
\item brightness-temperature $T_b$ image slices, for direct qualitative inspection;
\item the global signal $\langle T_b\rangle$ versus line-of-sight position or redshift, to test mean evolutionary behavior;
\item the power spectrum, to test two-point spatial structure;
\item reduced scattering coefficients, to test higher-order non-Gaussian morphology \citep{andreux2020kymatio,10.1093/mnras/staa3165,2023MNRAS.519.5288G}.
\end{enumerate}

% The global signal is the transversely averaged 21 cm brightness temperature as a function of line-of-sight position or redshift. In practice, we first average each lightcone over the two sky-plane dimensions and then aggregate the resulting curves across 800 realizations using the median. 
% % The power spectrum is computed from a fixed-redshift slice at \(z = 11.93\), corresponding to the central slice of the lightcone. 
% The power spectrum is computed from a square \(64\times64\) two-dimensional lightcone patch spanning one transverse direction and the line of sight, whose starting line-of-sight index corresponds to \(z = 11.93\).
% This redshift was chosen because the global signal, power spectrum, and scattering coefficients are well separated between the representative \((\log_{10}T_{\rm vir},\zeta)\) configurations at this stage, making qualitative differences easier to visualize and compare. The resulting statistics are likewise aggregated across realizations using the median.
% We use the same \(64\times64\) transverse--line-of-sight 2D lightcone patches
% to compute the reduced second-order scattering coefficients.
\rev{The global signal is the brightness temperature averaged over all
available transverse dimensions as a function of line-of-sight position or, equivalently,
redshift. For the 2D representation, this requires averaging over its single
transverse direction; for the 3D representation, we average over both
sky-plane dimensions. The resulting curves are then aggregated across 800
realizations using the median.}
\rev{For both dimensionalities, the power spectrum is evaluated on
$64\times64$ transverse--line-of-sight lightcone patches whose starting
line-of-sight index corresponds to $z=11.93$. For the 3D samples, we extract the central transverse--line-of-sight plane from each volume, while the same-sized region is taken directly from the native 2D output. This location was selected because the validation configurations exhibit more distinguishable signal evolution there. The resulting statistics are aggregated across realizations using the
median. We use the same patches to compute the reduced second-order scattering coefficients.}
The second-order scattering coefficients are constructed from cascaded wavelet convolutions followed by modulus operations and spatial averaging.
For an input 2D image \(I\), the second-order coefficients can be written as
\[
S_2(j_1,\ell_1,j_2,\ell_2)
=
\left\langle
\left|
\left| I \star \psi_{j_1,\ell_1} \right|
\star
\psi_{j_2,\ell_2}
\right|
\right\rangle ,
\]
where \(\psi_{j,\ell}\) denotes a wavelet filter at scale \(j\) and
orientation \(\ell\), \(\star\) denotes convolution, \(|\cdot|\) denotes the
modulus operation, and \(\langle\cdot\rangle\) denotes spatial averaging.
These coefficients probe correlations between image structures across
different spatial scales and orientations, and therefore provide a compact
description of multiscale non-Gaussian morphology.
% We use the same fixed-redshift 2D slices at \(z=11.93\) to compute reduced
% second-order scattering coefficients.
Following \citet{zhao2023can}, we
retain only scale pairs with \(j_2>j_1\), removing redundant scale
combinations while preserving the dominant structural information. In
practice, we average over the absolute orientation \(\ell_1\) and retain the
relative orientation class \((\ell_2-\ell_1)\bmod L\), which reduces the
number of coefficients while preserving orientation-dependent morphological
information. We additionally use logarithmic scattering coefficients for
dynamic-range compression, following conventions commonly adopted in prior
2D analyses \citep{zhao2023can}.

\begin{figure*}[t]
\centering
\includegraphics[width=0.49\textwidth]{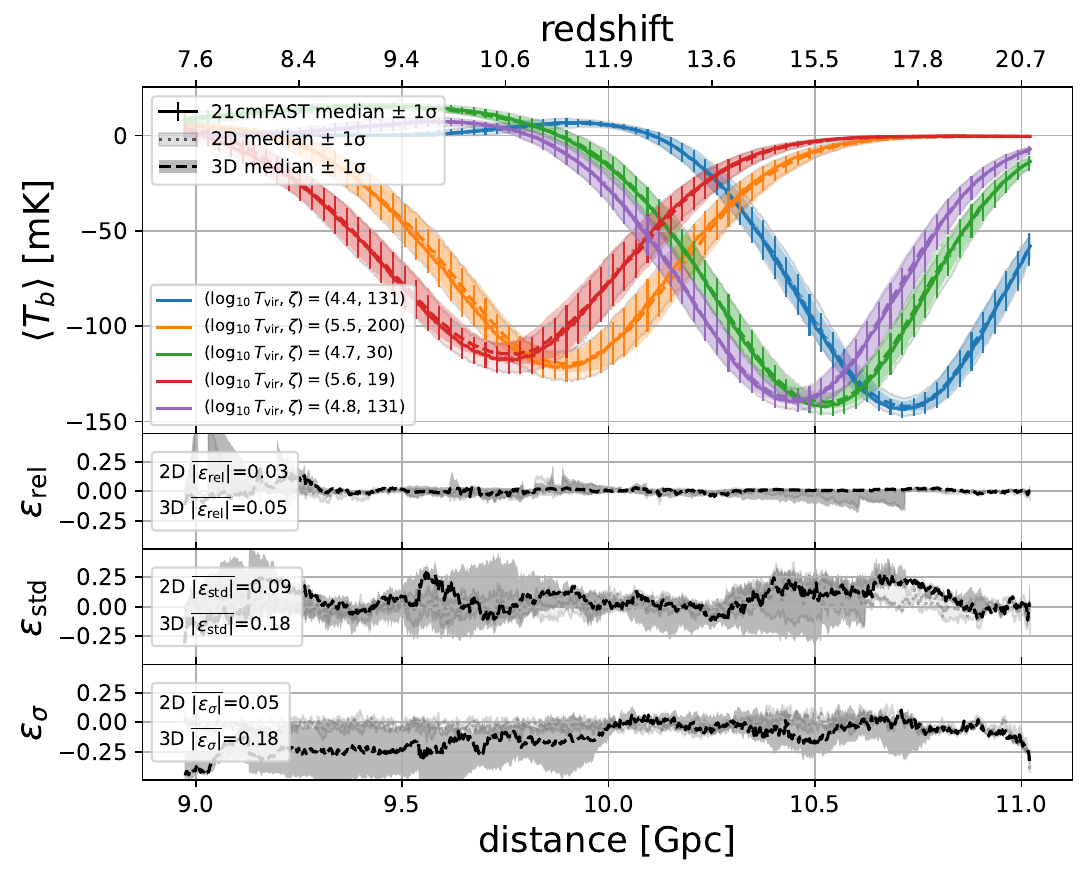}
\hfill
\includegraphics[width=0.49\textwidth]{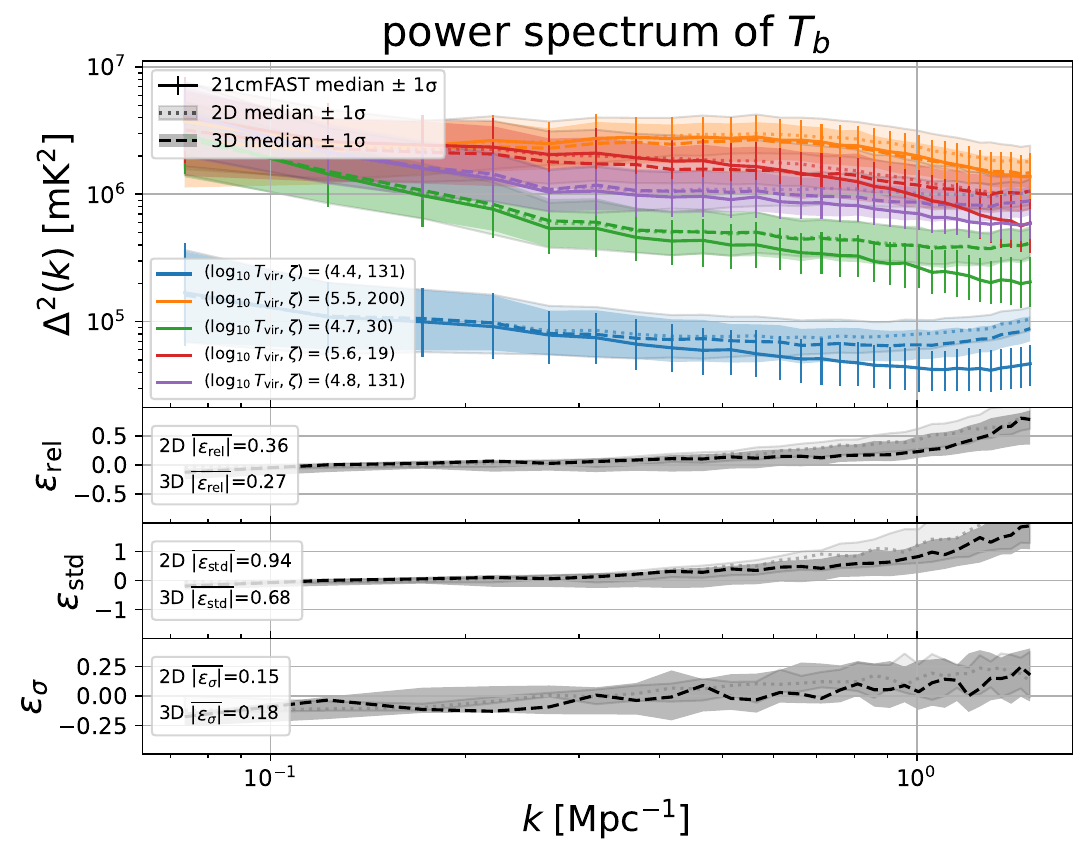}
\caption{
\rev{Global-signal (left) and power-spectrum (right) diagnostics for the
matched best 2D and 3D models, corresponding to indices 6 and 14 in
Table~\ref{tab:job_registry}. Both models use Yeo--Johnson preprocessing,
$A=0.1$, one residual block per encoder level, and 240 training epochs; the
3D model is the model visualized in Figure~\ref{fig:single_job_tb}. Colors
identify the five conditioning points, while line styles and markers
distinguish the \texttt{21cmFAST}, 2D, and 3D ensembles. 
In the upper panels, the error bars and shaded bands show the central
$68.27\%$ intervals across the 800 realizations at each validation parameter
point. In each residual panel, the line shows the median of the five
parameter-specific residual curves, while the shaded region shows their
central $68.27\%$ range. Each parameter-specific residual is computed after
reducing its 800-realization reference and generated ensembles to their
respective medians and scatter estimates.
\emph{Left:} median global
signal $\langle T_b\rangle$ as a function of distance/redshift and the
residual channels $\epsilon_{\rm rel}$, $\epsilon_{\rm std}$, and
$\epsilon_{\sigma}$. \emph{Right:} corresponding power-spectrum comparison
for the selected transverse--line-of-sight patches. The values printed in the
residual panels are the MAEs aggregated across all five validation points.}
}
\label{fig:single_job_global_power}
\end{figure*}

\rev{In the upper summary-statistic panels of
Figures~\ref{fig:single_job_global_power} and
\ref{fig:single_job_scattering}, the error bars and shaded bands show the
central $68.27\%$ intervals across the 800 realizations at each fixed
validation parameter point. In the corresponding residual panels, each
parameter point first yields a residual curve from the medians and scatter
estimates of its reference and generated ensembles. The plotted line is the
median of the five parameter-specific residual curves, and the shaded region
shows their central $68.27\%$ range. The shaded reference region in
Figure~\ref{fig:hparam_global_delta} instead represents the realization-level
scatter of the \texttt{21cmFAST} reference ensemble. These intervals visualize
ensemble or parameter-set variation and are not formal parameter-estimation
uncertainties.}

All major diagnostic functions share the same residual definitions:
\begin{align}
\epsilon_{\rm rel}
&=
\frac{y_1-y_0}{|y_0|},
\\
\epsilon_{\rm std}
&=
\frac{y_1-y_0}{\sigma_0},
\\
\epsilon_{\sigma}
&=
\frac{\sigma_1}{\sigma_0}-1,
\end{align}
where subscripts \(0\) and \(1\) denote the reference \texttt{21cmFAST} ensemble and the generated diffusion-model ensemble, respectively. 
\rev{Here \(y\) denotes the ensemble median, while \(\sigma\) denotes
one-half of the central \(68.27\%\) quantile interval, which we use as a
robust \(1\sigma\)-equivalent measure of ensemble scatter.}
We apply masking thresholds on \(|y_0|\) or \(\sigma_0\) to avoid artificial divergence when the denominator becomes too small.
\rev{These three channels probe distinct failure modes:
$\epsilon_{\rm rel}$ measures the relative bias with respect to the reference
signal, $\epsilon_{\rm std}$ measures the mismatch relative to the robust
ensemble scatter of the reference distribution, and $\epsilon_{\sigma}$
measures the fractional discrepancy in ensemble dispersion.}
The reported scalar metrics are the mean absolute errors (MAEs),
\begin{align}
{\rm MAE}_{\rm rel}
&=
\overline{|\epsilon_{\rm rel}|},
\\
{\rm MAE}_{\rm std}
&=
\overline{ |\epsilon_{\rm std}| },
\\
{\rm MAE}_{\sigma}
&=
\overline{ |\epsilon_{\sigma}| }.
\end{align}

% Unless otherwise stated, these metrics are computed from 800 generated samples and 800 reference 21cmFAST realizations for each validation parameter set.

Among these scalar summaries, we use $\mathrm{MAE}_{\rm std}$ as the primary compact ranking metric in the hyperparameter sweep. The reason is pragmatic as well as physical: $\mathrm{MAE}_{\rm rel}$ can become numerically unstable when the reference signal is close to zero, while $\mathrm{MAE}_{\sigma}$ is designed specifically to measure recovery of diversity rather than central accuracy. By normalizing the residual to the sample scatter of the reference ensemble, $\mathrm{MAE}_{\rm std}$ provides the most stable single-number measure of conditional bias across the diagnostics considered here. Unless explicitly noted otherwise, our ranking descriptions in the hyperparameter sweep refer to $\mathrm{MAE}_{\rm std}$ computed from the global-signal residual channels.

\section{Results}
\label{sec:results}

\subsection{\rev{Comparison of the Best 2D and 3D Models}}
\label{sec:what_is_right}

\begin{figure*}[t]
\centering
\includegraphics[width=\linewidth]{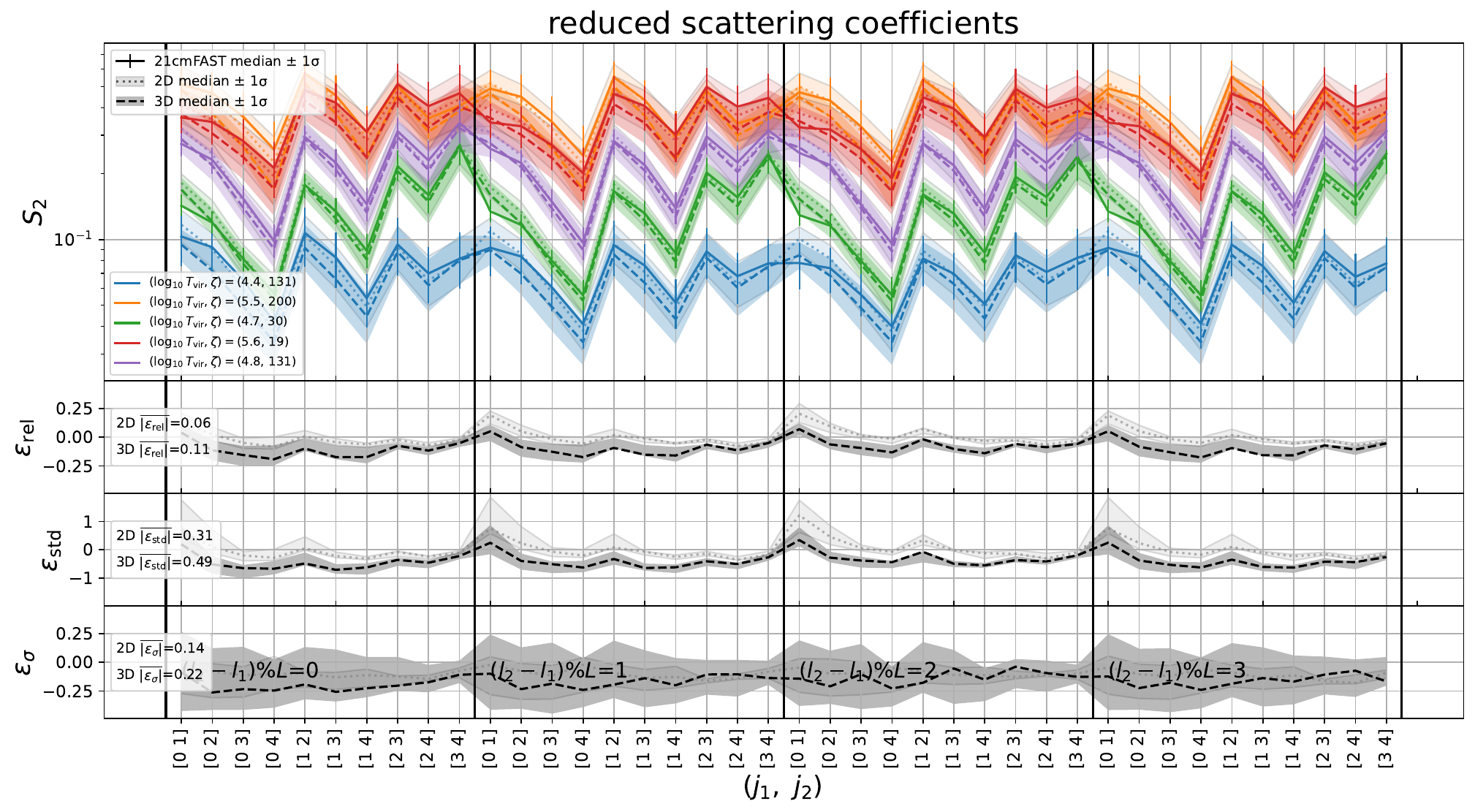}
\caption{
\rev{Reduced scattering-coefficient comparison for the matched best 2D and
3D models used in Figure~\ref{fig:single_job_global_power}. The top panel
compares the second-order coefficients $S_2$ across multiscale index pairs
$(j_1,j_2)$ and relative orientation classes, while the lower panels show the
corresponding residual channels. Colors identify the five conditioning
points, and the printed MAEs are aggregated across all five validation
points. In the upper panel, the error bars and shaded bands show the central
$68.27\%$ intervals across the 800 realizations at each validation point. In
the residual panels, the line and shaded region show the median and central
$68.27\%$ range, respectively, across the five parameter-specific residual
curves.}
}
\label{fig:single_job_scattering}
\end{figure*}

\rev{The full preprocessing and hyperparameter study is presented in
Appendix~\ref{app:hparam}. That analysis identifies the matched 240-epoch 2D
and 3D Yeo--Johnson models with $A=0.1$ and one residual block per encoder
level as the preferred models for the direct dimensional comparison. They
correspond to indices 6 and 14 in Table~\ref{tab:job_registry}, respectively.
At the fixed validation point, the lower global-signal
$\mathrm{MAE}_{\rm std}$ of the 2D model in the table indicates that the
lower-dimensional problem remains easier to optimize even after the 3D
training has been stabilized through preprocessing and extended training.
Figure~\ref{fig:single_job_tb} first illustrates the
volumetric output of the selected 3D model, while
Figures~\ref{fig:single_job_global_power}
and~\ref{fig:single_job_scattering} provide the matched quantitative
comparison.}

\rev{The 3D model captures the broad large-scale morphology and the principal
brightness-temperature evolution. In Figure~\ref{fig:single_job_tb}, it
reproduces the approximate positions and widths of the dominant transition
regions and generates visibly different realizations at fixed conditioning
parameters. Its most apparent residuals are local: the generated slices are
smoother in bubble-rich regions, do not always reproduce the sharpest
boundaries or finest structure, and contain small nonzero values within some
regions that are more cleanly concentrated near zero in the
\texttt{21cmFAST} samples. The purpose of Figure~\ref{fig:single_job_tb} is to
demonstrate the contiguous output of the 3D model; the relative accuracy of
the dimensional representations is assessed by the following ensemble
statistics.}

\rev{The left plot of Figure~\ref{fig:single_job_global_power} shows a clear
advantage for the 2D model in the global signal. The 2D emulator gives
$\mathrm{MAE}_{\rm rel}=0.03$, $\mathrm{MAE}_{\rm std}=0.09$, and
$\mathrm{MAE}_{\sigma}=0.05$, compared with $0.05$, $0.18$, and $0.18$ for
the 3D emulator. It therefore recovers both the median line-of-sight evolution
and the realization-to-realization dispersion more accurately. Both models
follow the broad timing and depth of the absorption troughs, while the
remaining errors are concentrated near the steepest transitions, where small
shifts in transition location generate visible residuals. Some residual
curves are truncated because bins with denominators below $1\,\mathrm{mK}$
are masked to avoid artificial divergence near zero crossings.}

\rev{The power-spectrum comparison is more nuanced. The 3D model has smaller
central-residual errors, with
$\mathrm{MAE}_{\rm rel}=0.27$ and
$\mathrm{MAE}_{\rm std}=0.68$, compared with 0.36 and 0.94 for the 2D model.
The 2D model gives a slightly smaller dispersion error,
$\mathrm{MAE}_{\sigma}=0.15$ rather than 0.18.
The 3D advantage in the central-residual channels is concentrated
primarily toward the high-$k$ end of the measured range. For
$k\gtrsim0.5\,{\rm Mpc}^{-1}$, the median 3D residuals in both
$\epsilon_{\rm rel}$ and $\epsilon_{\rm std}$ remain closer to zero than
their 2D counterparts, indicating a more accurate recovery of small-scale
power. At lower $k$, the two models give more comparable central residuals.
Thus, the additional spatial context available to the 3D model can improve
the recovery of some two-point statistics even though it does not produce a
uniform improvement. The comparison extends to wavenumbers of
$k\gtrsim1\,{\rm Mpc}^{-1}$, where both models exhibit scale-dependent
residuals and the smallest resolved structures remain difficult to reproduce.
This behavior is relevant for observational applications because foreground
avoidance and instrumental systematics can reduce access to the lowest-$k$
modes, increasing the importance of intermediate- and high-$k$ performance.
Related scale-dependent limitations have also been discussed in recent
field-level 21 cm emulation work \citep{2026JCAP...06..011P}.}
% \rev{The power-spectrum comparison is more nuanced. The 3D model has smaller
% central-residual errors, with $\mathrm{MAE}_{\rm rel}=0.27$ and
% $\mathrm{MAE}_{\rm std}=0.68$, compared with $0.36$ and $0.94$ for the 2D
% model. The 2D model gives a slightly smaller dispersion error,
% $\mathrm{MAE}_{\sigma}=0.15$ rather than $0.18$. Thus, the additional spatial
% context available to the 3D model can improve the recovery of some two-point
% statistics even though it does not produce a uniform advantage. The
% comparison extends to wavenumbers of $k\gtrsim1\,{\rm Mpc}^{-1}$, where both
% models exhibit scale-dependent residuals and the smallest resolved structures
% remain difficult to reproduce. This behavior is relevant for observational
% applications because foreground avoidance and instrumental systematics can
% reduce access to the lowest-$k$ modes, increasing the importance of
% intermediate- and high-$k$ performance. Related scale-dependent limitations
% have also been discussed in recent field-level 21 cm emulation work
% \citep{2026JCAP...06..011P}.}

\rev{Figure~\ref{fig:single_job_scattering} shows a second clear advantage for
the 2D representation. Its scattering-coefficient errors are
$\mathrm{MAE}_{\rm rel}=0.06$, $\mathrm{MAE}_{\rm std}=0.31$, and
$\mathrm{MAE}_{\sigma}=0.14$, compared with $0.11$, $0.49$, and $0.22$ for
the 3D model. Both models reproduce the broad ordering and oscillatory
structure of the coefficients across scale pairs and relative orientation
classes, but the 2D model more accurately recovers the non-Gaussian
morphology summarized by the reduced scattering transform. Higher-order
morphology therefore remains challenging in 3D even when the generated slices
appear visually plausible. 
The close agreement of the 2D scattering coefficients is qualitatively
consistent with \citet{zhao2023can}, who also found strong agreement in
scattering-based diagnostics for a conditional 2D DDPM. Their model was
trained on fixed-redshift transverse images, whereas our 2D baseline is
trained on transverse--line-of-sight lightcones, so the reported numerical
metrics are not directly comparable. Although our implementation also
supports training on fixed-redshift images, and fixed-redshift transverse
slices can be extracted from the 3D outputs, a controlled comparison would
require matched target geometry, redshift selection, training data,
preprocessing, and evaluation metrics.}

% \rev{A complementary one-point comparison is provided by the voxel PDFs in
% Figure~\ref{fig:hparam_pdf}. The best 2D and 3D models have very similar
% transformed- and physical-space voxel distributions and share the same
% principal deficiency: both broaden the sharply concentrated
% $T_b\simeq0$ component of the reference distribution. The one-point PDF
% therefore does not provide a meaningful dimensional advantage and is retained
% in Appendix~\ref{app:hparam}, where it also serves its primary purpose of
% diagnosing preprocessing choices. The more substantial differences between
% the 2D and 3D models appear in the line-of-sight and spatial statistics shown
% in Figures~\ref{fig:single_job_global_power}
% and~\ref{fig:single_job_scattering}.}
\rev{A complementary one-point comparison is provided by the voxel PDFs in Figure~\ref{fig:hparam_pdf} for the matched 120-epoch 2D and 3D models (indices 5 and 15 in Table~\ref{tab:job_registry}). Their transformed- and physical-space voxel distributions are very similar and share the same principal deficiency: both broaden the sharply concentrated $T_b \simeq 0$ component of the reference distribution. At this training duration, the one-point PDF therefore does not provide a meaningful dimensional advantage and is retained in Appendix~\ref{app:hparam}, where it also serves its primary purpose of diagnosing preprocessing choices. The more substantial differences between the 2D and 3D models appear in the line-of-sight and spatial statistics shown in Figures~\ref{fig:single_job_global_power} and~\ref{fig:single_job_scattering}.}

\rev{Taken together, these results show that increased dimensionality is a
major optimization challenge. Under the architecture and computational budget
explored here, the additional information in the 3D volume does not
consistently compensate for the larger output space, smaller micro-batches,
and more difficult optimization: the 2D model performs better for the global
signal and scattering coefficients and recovers the ensemble dispersion more
accurately in all three plotted diagnostics. The conclusion is not completely
one-sided, however, because the 3D model gives smaller central-bias errors in
the power spectrum. We therefore characterize the result as a
dimensionality--fidelity trade-off rather than as a universal superiority of
either representation.}

\section{Discussion}
\label{sec:discussion}
\subsection{\rev{The Dimensionality--Fidelity Trade-off}}

\rev{The matched comparison demonstrates that increasing the information
content of the target does not automatically increase emulator accuracy. 
Both matched representations retain the line-of-sight evolution, while
the 3D lightcone additionally contains correlations across the second
transverse direction. Learning this larger conditional distribution requires
substantially more memory, smaller micro-batches, and noisier optimization.
Under the present architecture and training budget, these costs
outweigh the additional information for the global signal and scattering
coefficients, while the 3D representation retains an advantage in the central
power-spectrum residuals, particularly toward the high-$k$ end of the
measured range.}

\rev{These results suggest that a 2D transverse--line-of-sight representation
may be the more practical training space when the primary objective is
computationally efficient recovery of selected summary statistics. It is not,
however, a substitute for a 3D emulator when the scientific application
requires contiguous volumetric realizations, correlations across both
transverse directions, or statistics that cannot be reconstructed from
independent 2D outputs. The present 3D model should therefore be viewed as a
volumetric baseline rather than as a uniformly more accurate replacement for
the 2D emulator.}
\subsection{Challenges in 3D Lightcone Emulation}
The move from 2D image generation to 3D lightcone emulation is not simply a dimensional upgrade. The dominant failure modes in our experiments arise from coupled computational and statistical effects: memory constraints force small micro-batches, small batches amplify optimization noise, and unstable normalization magnifies the resulting errors. In this regime, preprocessing is not a minor implementation detail but a principal modeling choice.

Our results also show that evaluation must be physically layered. The global signal, power spectrum, and scattering coefficients probe distinct aspects of fidelity: mean evolution, two-point structure, and higher-order non-Gaussian morphology. This hierarchy matters for 21 cm lightcones because their information content is not fully described by Gaussian statistics. A model may therefore appear satisfactory under visual inspection or even under low-order summaries while still retaining meaningful biases in morphology-sensitive diagnostics. That is precisely why the scattering transform is useful here: it helps expose discrepancies that would be missed by image inspection or by the global signal alone, consistent with the broader conclusions of \citet{zhao2023can}. \rev{More broadly, our results support combining image-space comparisons, lower-order summaries, and morphology-sensitive diagnostics when the scientific goal depends on preserving non-Gaussian structure.}

\rev{The validation design also permits a direct comparison of conditional
diversity. Because the reference sets contain many independent realizations at
fixed parameter points, they test whether each model learns a distribution
rather than only a deterministic mean mapping. Both models generate
realization-to-realization variation, but the 2D model gives smaller
$\mathrm{MAE}_{\sigma}$ values for the global signal, power spectrum, and
scattering coefficients. The greater difficulty of the 3D problem therefore
affects not only the conditional median but also the detailed recovery of the
ensemble dispersion. This remains important for inference-oriented
applications, for which realistic sample diversity is essential.}

\subsection{Limitations and Future Work}
This paper focuses on emulator fidelity under simulation-level conditions rather than a complete survey-facing inference pipeline. Several ingredients remain necessary for that next step: instrument response and foreground contamination, uncertainty calibration beyond percentile envelopes, larger controlled studies of realization-to-realization variance, and direct downstream validation in simulation-based inference (SBI) or Markov chain Monte Carlo (MCMC) workflows. One specific open question is whether the small-scale smoothing \rev{identified in Section~\ref{sec:what_is_right} and Appendix~\ref{sec:what_still_fails}} would lead to measurable parameter-level bias in downstream inference. Quantifying that effect will require explicit SBI or MCMC tests and is beyond the scope of this paper. 
\rev{A further controlled scaling study will be needed to determine whether
the remaining 2D--3D gap is primarily optimization-limited and can be reduced
through larger effective batches, alternative 3D architectures, or increased
training compute, or whether a different target representation is required.}
The present work should therefore be viewed as a baseline study of generative fidelity rather than as a complete end-to-end inference demonstration.

The present validation is also interpolative: all reference parameter points lie within the training range. In addition, the training data use one lightcone per training parameter pair, whereas the validation ensembles use many realizations at fixed parameter points. This design is efficient for parameter-space coverage, but future work could explore denser multi-realization training sets, wider astrophysical conditioning spaces, and more explicit tests of generalization under distribution shift. Finally, although the present analysis uses 800-sample ensembles for both the generated and reference sets, more extensive sampling studies could still be valuable for tightening uncertainty estimates.

More broadly, the main astrophysical value of this work lies in methodology and benchmarking. Before a 3D generative emulator can be trusted for survey-facing 21 cm inference, it must be validated in a way that is sensitive not only to image appearance but also to the physically meaningful statistics of the signal. The present study identifies preprocessing and diagnostic choices that materially affect that trustworthiness and therefore helps define a more robust baseline for future emulator development.

\section{Conclusion}
\label{sec:conclusion}

\rev{We have presented a controlled study of conditional diffusion modeling
for 3D 21 cm lightcone emulation together with a matched 2D baseline. In this
high-dimensional setting, stable and physically meaningful generation depends
strongly on preprocessing and dynamic-range control. Among the configurations
tested here, Yeo--Johnson preprocessing with moderate amplitude compression
provides the most consistently favorable trade-off, while increasing
architecture depth does not produce a monotonic improvement.}

\rev{The matched dimensional comparison shows that extending the target from
2D to 3D does not automatically improve physical fidelity. The 2D model gives
smaller global-signal and scattering-coefficient errors and more accurately
recovers the ensemble dispersion across all three plotted diagnostics. 
The 3D model nevertheless gives smaller central-residual errors in the power
spectrum, particularly at high $k$, indicating that it can exploit part of
the additional spatial information to improve the recovery of small-scale
power.
The result is therefore diagnostic dependent: the increased
dimensionality is a larger challenge than the added information can
consistently compensate for under the present training setup, but the 3D
representation remains necessary for applications requiring contiguous
volumetric samples and full transverse coherence.}

\rev{Across both representations, the global signal is recovered more
accurately than the detailed spatial statistics. The broadened near-zero voxel
component, smoothing of low-$T_b$ regions, and scale-dependent residuals at
high $k$ show that visually convincing samples can still retain physically
meaningful biases. The present work therefore provides a reproducible
simulation-level baseline for evaluating 21 cm emulators through complementary
image-space, one-point, two-point, and morphology-sensitive diagnostics.
Future work should determine whether improved 3D optimization and architecture
design can convert its larger information content into a more uniform gain in
emulator fidelity.}

\begin{acknowledgments}
We acknowledge funding support from a Georgia Tech Research Institute IRAD award, NSF grants AST-2108020 and AST-2510197, and NASA grant 80NSSC21K1053. This research used resources of the National Energy Research Scientific Computing Center (NERSC), a U.S. Department of Energy Office of Science User Facility, through the GenAI@NERSC award.
\end{acknowledgments}

% \section*{data availability}
% The training and validation data underlying this article are available from the corresponding author upon request. The code used for the analysis and model development is publicly available at \url{https://github.com/Xsmos/ml21cm}.
\section*{Data and Software Availability}

The training and validation data underlying this article are available from
the corresponding author upon reasonable request.
\rev{The software used for data generation, model training, sampling, and
analysis is available on GitHub and has been archived as version 1.0.0 on
Zenodo \citep{Xia2026ml21cm}.}

\software{
\texttt{21cmFAST} \citep{Mesinger_2007,2011MNRAS.411..955M,Murray2020},
PyTorch \citep{paszke2019pytorch},
NumPy \citep{2020Natur.585..357H},
Matplotlib \citep{Hunter:2007},
\rev{\texttt{ml21cm} \citep{Xia2026ml21cm}}
}
%% Appendix material should be preceded with a single \appendix command.
%% There should be a \section command for each appendix. Mark appendix
%% subsections with the same markup you use in the main body of the paper.
%%
%% Each Appendix (indicated with \section) will be lettered A, B, C, etc.
%% The equation counter will reset when it encounters the \appendix
%% command and will number appendix equations (A1), (A2), etc. The
%% Figure and Table counter will not reset.

\appendix

\twocolumngrid

\section{Hyperparameter Sweep}
\label{app:hparam}
\subsection{Model Performance Dependence on Preprocessing}

\begin{figure*}[t]
\centering
\includegraphics[width=\linewidth]{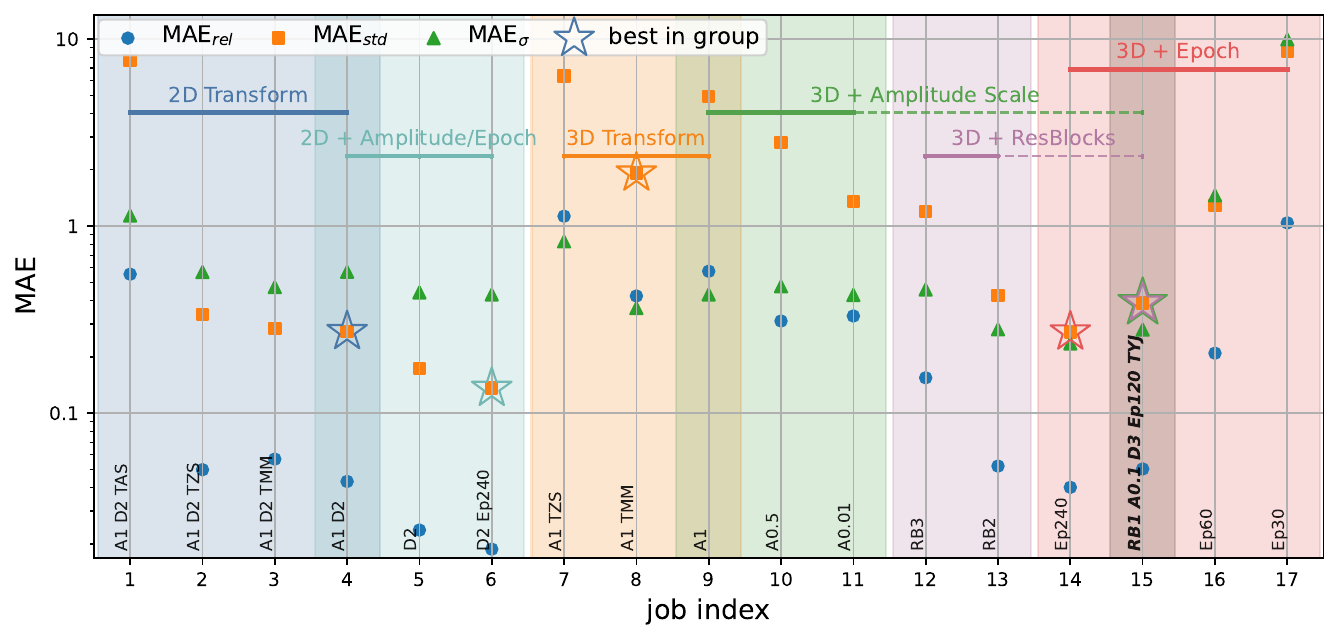}
\caption{
Summary of scalar MAE metrics across the hyperparameter sweep. Markers show $\mathrm{MAE}_{\rm rel}$, $\mathrm{MAE}_{\rm std}$, and $\mathrm{MAE}_{\sigma}$ for each indexed run listed in Table~\ref{tab:job_registry} where their labels are given; open stars mark the best run within each comparison group using $\mathrm{MAE}_{\rm std}$ as the primary selection metric. The colored guide bands separate the 2D-transform tests, the matched 2D amplitude/training-duration tests, the 3D-transform tests, the 3D amplitude-scaling tests, the residual-block tests, and the 3D epoch-length tests. All metrics in this figure are evaluated at
$(\log_{10}T_{\rm vir},\zeta)=(4.699,30)$.
}
\label{fig:hparam_mae}
\end{figure*}

\begin{figure}[t]
\centering
\includegraphics[width=\linewidth]{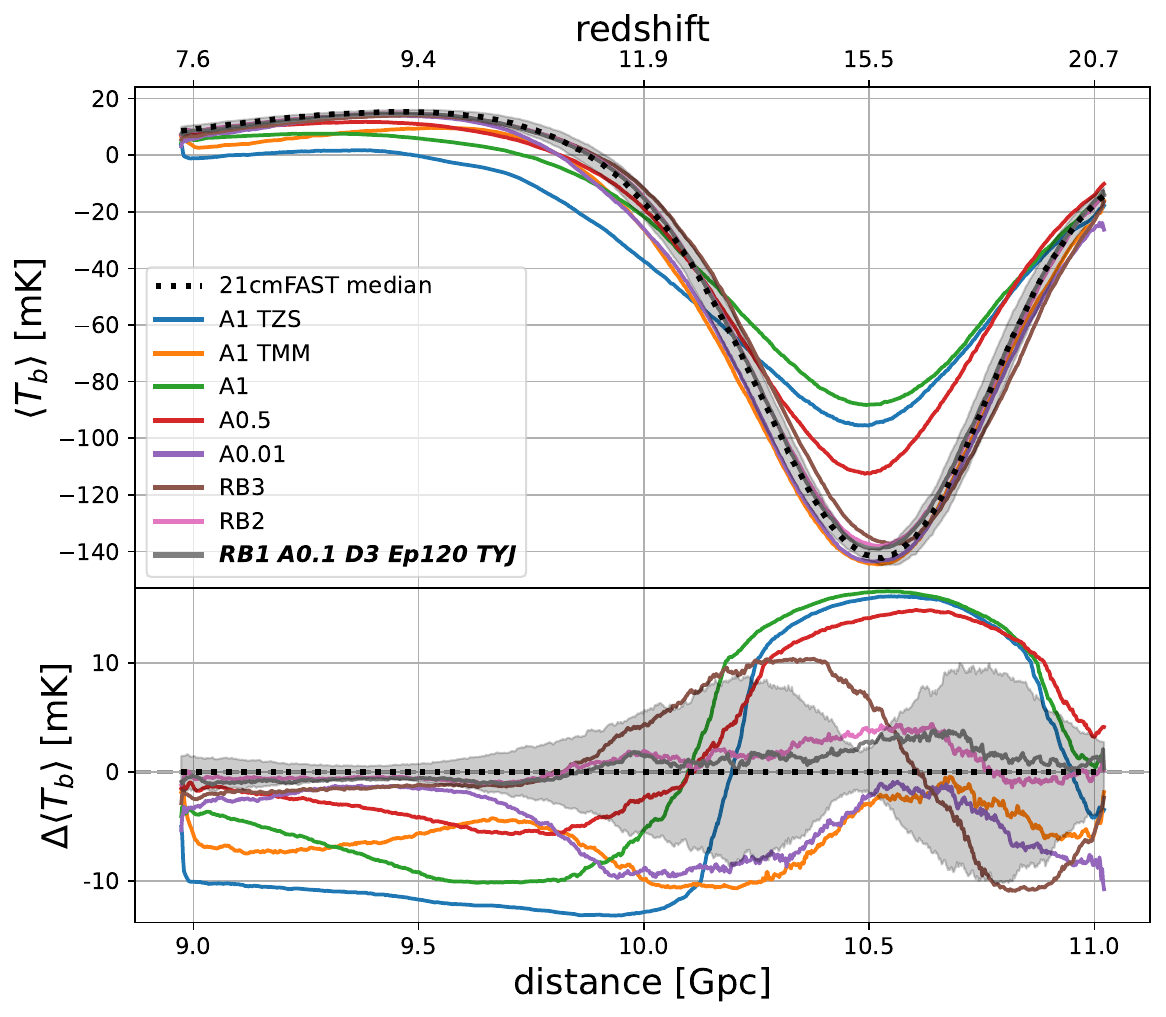}
\caption{
Comparison of selected 3D runs against \texttt{21cmFAST} (dotted line) using the global signal and its residual, $\Delta\langle T_b\rangle$. The upper panel shows the median $\langle T_b\rangle$ evolution for representative runs spanning different preprocessing choices, amplitude scalings, and residual-block depths, and the lower panel shows the corresponding deviation from the \texttt{21cmFAST} median. This comparison is shown at $(\log_{10}T_{\rm vir},\zeta)=$ $(4.699,30)$. The gray band denotes the central 68.27\% interval across the 800 21cmFAST realizations.
}
\label{fig:hparam_global_delta}
\end{figure}

\begin{table*}[!htbp]
\centering
\caption{
\rev{Index-to-hyperparameter mapping used in the hyperparameter-comparison
figures. The fiducial 3D configuration is highlighted in bold and serves as
the comparison anchor in the sweep. The best 3D model visualized in
Figure~\ref{fig:single_job_tb} corresponds to index 14, while the matched 2D
and 3D models compared in Figures~\ref{fig:single_job_global_power}
and~\ref{fig:single_job_scattering} correspond to indices 6 and 14,
respectively. The \textit{Label} column gives the shorthand labels used in
Figures~\ref{fig:hparam_mae} and~\ref{fig:hparam_global_delta}. 
The final column reports the global-signal
$\mathrm{MAE}_{\rm std}$ evaluated at the fixed validation point
$(\log_{10}T_{\rm vir},\zeta)=(4.699,30)$ for every run. The main-text MAEs
for the matched models are instead evaluated over all five validation points.
% The main-text MAEs for the matched models are instead evaluated over all five validation points and therefore need not be numerically identical to the values reported here.
}
}
\label{tab:job_registry}
\begin{tabular*}{\textwidth}{@{\extracolsep{\fill}}cccccccc@{}}
\toprule
Index & Dim & Transform & Amplitude & ResBlocks & Epochs & Label
      & MAE$_{\rm std}$ \\
\midrule
\multicolumn{8}{c}{\textit{2D runs}} \\
\midrule
1 & 2D & arcsinh
  & 1   & 1 & 120 & A1 D2 TAS          & 7.67  \\
2 & 2D & z-score
  & 1   & 1 & 120 & A1 D2 TZS          & 0.338 \\
3 & 2D & min--max
  & 1   & 1 & 120 & A1 D2 TMM          & 0.284 \\
4 & 2D & Yeo--Johnson
  & 1   & 1 & 120 & A1 D2              & 0.274 \\
5 & 2D & Yeo--Johnson
  & 0.1 & 1 & 120 & A0.1 D2            & 0.174 \\
6 & 2D & Yeo--Johnson
  & 0.1 & 1 & 240 & A0.1 D2 Ep240      & 0.136 \\
\midrule
\multicolumn{8}{c}{\textit{3D runs}} \\
\midrule
7  & 3D & z-score
   & 1    & 1 & 120 & A1 TZS       & 6.35  \\
8  & 3D & min--max
   & 1    & 1 & 120 & A1 TMM       & 1.93  \\
9  & 3D & Yeo--Johnson
   & 1    & 1 & 120 & A1           & 4.94  \\
10 & 3D & Yeo--Johnson
   & 0.5  & 1 & 120 & A0.5         & 2.80  \\
11 & 3D & Yeo--Johnson
   & 0.01 & 1 & 120 & A0.01        & 1.36  \\
12 & 3D & Yeo--Johnson
   & 0.1  & 3 & 120 & RB3          & 1.20  \\
13 & 3D & Yeo--Johnson
   & 0.1  & 2 & 120 & RB2          & 0.425 \\
14 & 3D & Yeo--Johnson
   & 0.1  & 1 & 240 & Ep240        & 0.272 \\
\textbf{15}
   & \textbf{3D}
   & \textbf{Yeo--Johnson}
   & \textbf{0.1}
   & \textbf{1}
   & \textbf{120}
   & \textbf{FIDUCIAL}
   & \textbf{0.389} \\
16 & 3D & Yeo--Johnson
   & 0.1  & 1 & 60  & Ep60         & 1.29  \\
17 & 3D & Yeo--Johnson
   & 0.1  & 1 & 30  & Ep30         & 8.63  \\
\bottomrule
\end{tabular*}
\end{table*}
% \FloatBarrier
% We begin with the different preprocessing methods listed in Section \ref{subsec:normalization} and the hyperparameter sweep because it provides the clearest quantitative view of which design choices have the largest impact on accuracy. Table~\ref{tab:job_registry} lists the indexed runs used in the comparison figures. The fiducial configuration, highlighted in bold, serves as the comparison anchor for the 3D sweep. 
% Figure~\ref{fig:hparam_mae} summarizes the preprocessing and hyperparameter sweep across the 2D and 3D runs. Figure~\ref{fig:hparam_global_delta} illustrates how these differences appear directly in the global-signal evolution. 
% The representative model shown later in Figures~\ref{fig:single_job_tb}--\ref{fig:single_job_scattering} is the corresponding 240-epoch extension of that fiducial setup.
\rev{We present the full preprocessing and hyperparameter sweep in this
Appendix because it provides the quantitative basis for selecting the models
used in the main dimensional comparison. Table~\ref{tab:job_registry} lists
the indexed runs. Figure~\ref{fig:hparam_mae} summarizes the scalar residual
metrics across the 2D and 3D configurations,
Figure~\ref{fig:hparam_global_delta} shows how the 3D differences appear in
the global-signal evolution, and Figure~\ref{fig:hparam_pdf} compares the
voxel distributions of the matched 120-epoch 2D and 3D models
(indices 5 and 15 in Table~\ref{tab:job_registry}). The best 3D model shown in
Figure~\ref{fig:single_job_tb} and the matched models compared in
Figures~\ref{fig:single_job_global_power}
and~\ref{fig:single_job_scattering} are the 240-epoch extensions identified
in Table~\ref{tab:job_registry}.}

\rev{Table~\ref{tab:job_registry} and
Figures~\ref{fig:hparam_mae}--\ref{fig:hparam_global_delta} show that
preprocessing and dynamic-range control are dominant drivers of performance,
particularly in 3D. The 2D runs serve both as lower-dimensional controls for
the preprocessing choices and as the matched baseline used in the main
scientific comparison. Within the transform-only tests, Yeo--Johnson and
min--max give the strongest 2D performance. Applying the corresponding
transforms in 3D produces global-signal $\mathrm{MAE}_{\rm std}$ values that
are approximately one order of magnitude larger, demonstrating the
dimensional penalty before additional compression is introduced.}

\rev{Within the 3D transform-only subset, plain z-score and plain
Yeo--Johnson perform markedly worse than min--max. This ordering changes once
moderate amplitude compression is added to the Yeo--Johnson transform. The
important distinction is not that the marginal 3D voxel distribution is
intrinsically broader than its 2D counterpart---the voxel PDFs in
Figure~\ref{fig:hparam_pdf} are in fact similar---but that the larger 3D
output space and the small micro-batches make the optimization more sensitive
to the same skewed and long-tailed distribution. Compressing the transformed
dynamic range therefore has a much larger practical impact on the 3D
training.}

\rev{The amplitude sweep shows that the improvement is not monotonic with
compression strength. The intermediate value $A=0.1$ performs better than
$A=1$, $0.5$, or $0.01$ and surpasses the plain min--max baseline. With
$A=0.1$ and extended 240-epoch training, the global-signal
$\mathrm{MAE}_{\rm std}$ gap between the matched 2D and 3D models is reduced
to approximately a factor of two, but it is not eliminated.}

Figure~\ref{fig:hparam_global_delta} depicts how these scalar differences appear in the global-signal $\langle T_b \rangle$ evolution itself: preprocessing and dynamic-range control primarily determine whether the absorption trough is placed at the correct depth and redshift, whereas architectural changes such as residual-block depth modulate the remaining mismatch more subtly. In particular, the better-performing compressed Yeo--Johnson runs place the absorption trough closer to the \texttt{21cmFAST} reference signal in both depth and redshift and remain closest to the reference curve over most of the lightcone, while several alternative settings produce systematic offsets that persist across extended redshift intervals.

The remaining trends are secondary. Figure~\ref{fig:hparam_mae} shows that increasing residual-block depth is not monotonically beneficial: the run with one residual block (RB1) outperforms the model with two (RB2) and does markedly better than the model with three (RB3), despite RB3 requiring far greater computational cost (1061-GPU hours for RB3, 536 GPU-hours for RB2, and 347-GPU hours for RB1). This result indicates that simply stacking more parameters does not guarantee better performance for this problem. One plausible interpretation is that, in the small-micro-batch regime enforced by 3D memory limits, the potential representational benefit of additional depth is outweighed by noisier optimization dynamics, so deeper models are not rewarded unless training is already sufficiently stable. 

The points in the red shaded region of Figure~\ref{fig:hparam_mae} show a consistent improvement from 30 to 240 epochs, with the 240-epoch run giving the best result among the tested values, although longer training runs were not explored. 
% Owing to computing time and memory constraints, the runs on the left side of the sweep were all trained for 120 epochs. 
Most transform, amplitude, and residual-block comparisons were trained for
120 epochs, while indices 6 and 14 are explicit 240-epoch extensions and
indices 16 and 17 test shorter training durations.
The representative example used in Figures~\ref{fig:single_job_tb}--\ref{fig:single_job_scattering} corresponds to the 240-epoch configuration. Taken together, our hyperparameter sweep identifies a consistent trend among
the tested configurations: the Yeo--Johnson preprocessing transform combined with a moderate amplitude compression gives the most consistently favorable trade-off between performance and compute time and memory usage, with the strongest direct support coming from the global-signal $\mathrm{MAE}_{\rm std}$ ranking.

\subsection{\rev{Voxel-Distribution Diagnostics and Remaining Inaccuracies}}
\label{sec:what_still_fails}
\begin{figure*}[t]
\centering
\includegraphics[width=\linewidth]{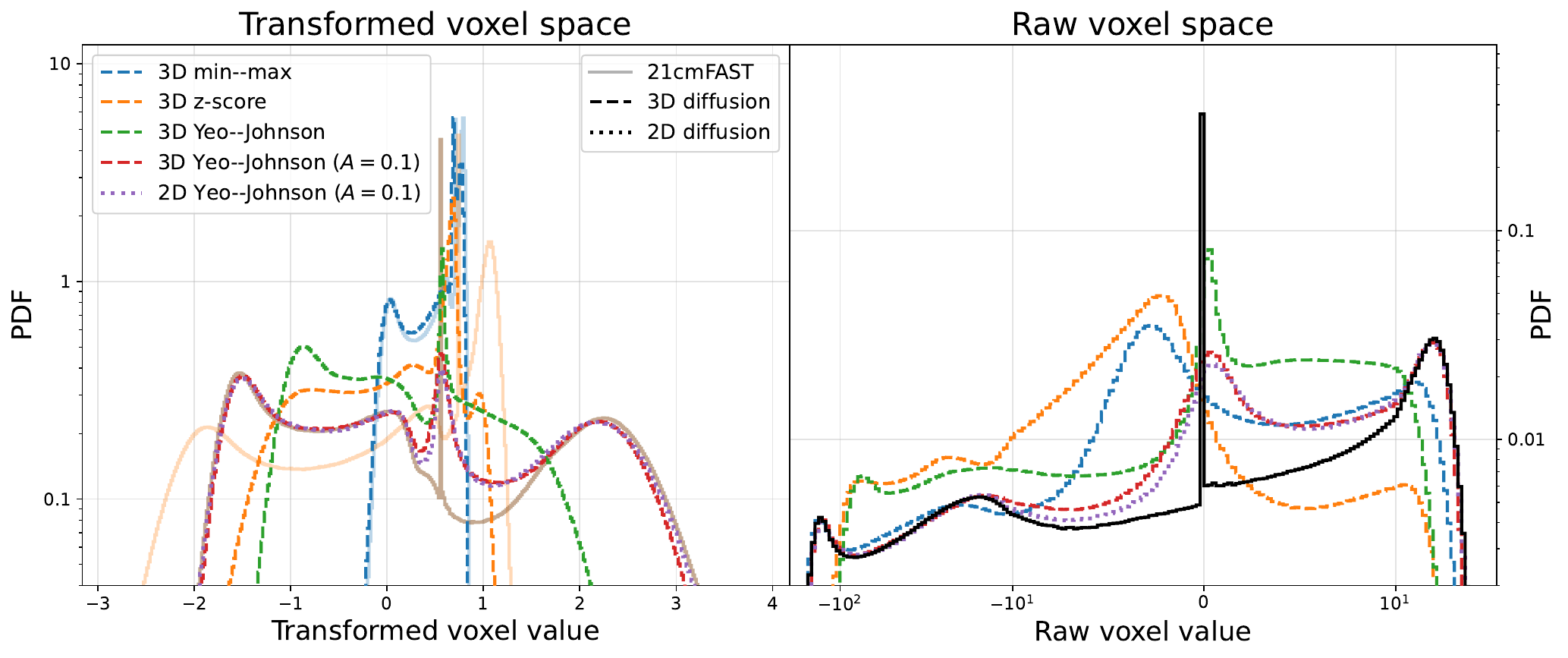}
\caption{
\rev{Comparison of preprocessing choices using lightcone voxel-value
distributions. The panels show probability density functions in transformed
space (left) and raw brightness-temperature space (right) for
\texttt{21cmFAST} targets and diffusion-generated samples. The matched
120-epoch 2D and 3D Yeo--Johnson models with $A=0.1$
(indices 5 and 15 in Table~\ref{tab:job_registry}) are included together with
the alternative 3D preprocessing configurations. All distributions are
evaluated at $(\log_{10}T_{\rm vir},\zeta)=(4.699,30)$.}
}
\label{fig:hparam_pdf}
\end{figure*}
% To isolate the remaining failure modes more directly, Figure~\ref{fig:hparam_pdf} compares voxel value probability density functions in both transformed space and raw space, which shows that an important part of the difficulty already appears in transformed voxel space. The two panels show voxel value probability density functions in transformed space (left) and raw brightness temperature $T_b$ space (right) for \texttt{21cmFAST} targets and diffusion-generated samples, respectively. The main challenge is the strong concentration of voxels near $0\,\mathrm{mK}$ in the physical lightcone distribution: the reference data contain a sharply peaked, nearly delta-function-like near-zero component, whereas the diffusion model reproduces this feature only as a peak with finite width.
% Although some preprocessing choices place the dominant near-zero peak
% at roughly the correct location, the diffusion model does not fully recover
% its height, width, or surrounding distribution.
\rev{Figure~\ref{fig:hparam_pdf} diagnoses the voxel-level effects of the
preprocessing choices in transformed and physical brightness-temperature
space. The matched 120-epoch 2D and 3D distributions are very similar over
most of their support and nearly overlap in both panels. The voxel PDF
therefore does not provide a strong distinction between the dimensionalities.
Its primary value is instead to identify a failure mode shared by both models
and to explain why additional dynamic-range compression improves the 3D
training.}

\rev{The main shared difficulty is the strong concentration of physical
voxels near $0\,\mathrm{mK}$. The \texttt{21cmFAST} distribution contains a
sharply peaked, nearly delta-function-like component, whereas both diffusion
models broaden it into a peak with finite width. Although some preprocessing
choices place the peak at approximately the correct location, they do not
fully recover its height, width, or surrounding distribution.}

\rev{This failure mode suggests a possible future improvement. A model with an auxiliary output head predicting whether a voxel belongs to the near-zero component could separate the nearly discrete near-zero population from the smoother nonzero brightness-temperature distribution. The diffusion model would then only need to model the continuous component conditional on this mask, which may reduce the artificial broadening of the delta-like peak near \(T_b \simeq 0\). We leave such a two-component architecture to future work.}
In transformed voxel space, min--max performs better than plain z-score and plain Yeo--Johnson because it compresses the data into a narrower range, making the target distribution easier for the diffusion model to reproduce. This observation helps explain why a secondary squish step can improve 3D training: it further narrows the transformed dynamic range that the diffusion model must learn.

The same figure also clarifies why transformed-space improvement is not the whole story. A Yeo--Johnson transform followed by an additional amplitude compression with $A=0.1$ yields the best overall agreement in both transformed and raw space, substantially improving upon the case without the additional compression. After mapping the samples back to physical brightness-temperature space, however, the near-zero component remains visibly broader than in the \texttt{21cmFAST} targets, indicating that the model still does not perfectly match this part of the distribution. In the \texttt{21cmFAST} reference, the near-zero voxel population is sharply concentrated and nearly delta function-like, whereas the diffusion model broadens it into a finite-width distribution. After inversion back to raw brightness-temperature space, voxel values that should remain tightly clustered near $0$~mK are instead spread over a continuous range, including small positive and negative values. In image space, this broadening manifests as visually dark regions that are less cleanly separated from their surroundings, with small nonzero pixels appearing inside regions that are closer to zero in the reference maps. This interpretation is consistent with the local residuals seen in Figure~\ref{fig:single_job_tb}, and it also provides a plausible link to the residual power-spectrum mismatch at high $k$: once low $T_b$ interiors and sharp boundaries are smeared into a broader continuum, the resulting fields naturally lose some of the smallest-scale contrast even if the large-scale trend is preserved.

Together with Figure~\ref{fig:hparam_global_delta}, the voxel-PDF comparison also shows why a voxel-level agreement alone is not sufficient to identify a preferred preprocessing strategy. Min--max is relatively competitive in the voxel-PDF comparison, but its global-signal behavior remains systematically biased, with the median brightness-temperature curve consistently shifted below the \texttt{21cmFAST} reference over much of the lightcone evolution. This behavior is consistent with the strong skewness of the voxel distribution: although min--max scaling preserves the overall range, it does not sufficiently regularize the highly asymmetric near-zero population, leading the diffusion model to underestimate the brightness temperature after inversion back to physical units.
In contrast, the better compressed Yeo--Johnson runs yield a more favorable compromise across voxel level and global signal diagnostics: the reconstructed median fluctuates around the \texttt{21cmFAST} reference while remaining largely within the reference \(1\sigma\) band throughout most of the lightcone evolution. Overall, Figure~\ref{fig:hparam_pdf} shows that preprocessing plays a major role in voxel-level fidelity: additional dynamic-range compression improves the agreement, but does not fully eliminate the difficulty of reproducing the sharply concentrated near-zero voxel population. The preferred configuration should therefore be understood not as universally optimal across every metric, but as the most consistent trade-off among the tested configurations.

\section{Implementation Details for Reproducibility}

\subsection{Architecture and Diffusion Settings}
The main architectural configuration is summarized in Section~3.2. Here we record only the implementation details most relevant for reproducibility.

In \texttt{ContextUnet}, 2D and 3D are unified through
dimension-dependent convolution and resampling operators. The residual block
uses GroupNorm with 32 groups, a SiLU activation, dimension-appropriate
$3\times3$ or $3\times3\times3$ convolutions, timestep-conditioned embedding
injection, dropout, and a final zero-initialized convolution, together with a
residual skip connection. Depending on the configuration, the skip path is
the identity or uses a dimension-appropriate pointwise or $3^d$ convolution.
When upsampling or downsampling is requested inside the residual block, the
feature path and skip path are resampled in parallel before the final
convolution.

The attention block normalizes the feature tensor, projects it to query, key,
and value tensors using a pointwise $1\times1$ convolution in 2D or
$1\times1\times1$ convolution in 3D, applies scaled dot-product attention in
flattened spatial coordinates, and returns the result through a residual
connection.

In \texttt{training/diffusion.py} provided in our GitHub repository, the DDPM training uses the following settings:
\begin{enumerate}
\item \texttt{betas=(1e-4, 0.02)} with a linear schedule,
\item \texttt{num\_timesteps=1000},
\item MSE loss on predicted noise,
\item AdamW optimizer with an initial learning rate of $10^{-5}$,
\item cosine annealing learning-rate schedule, and
\item gradient clipping (\texttt{max\_norm=1.0}).
\end{enumerate}
We also tested an exponential moving average (EMA) with a decay rate $\beta=0.995$ but found no improvement in the final comparisons, therefore the reported results do not use EMA.

\subsection{Data Pipeline and Normalization}
For each sample, the loader returns an image tensor and a normalized condition vector:
\begin{equation}
(\mathbf{x},\mathbf{c}),\quad \mathbf{c}\in[0,1]^2,
\end{equation}
where $\mathbf{c}$ corresponds to $(\log_{10}T_{\rm vir},\zeta)$ after min--max rescaling. Let $\mathbf{x}$ denote a raw voxel value of the 21 cm brightness-temperature field. Image-space preprocessing is selected via \texttt{scale\_path} and uses the four normalization choices listed in Section~\ref{subsec:normalization}. After the primary normalization step, the training code can additionally apply the squish transform
\begin{equation}
x_{\mathrm{sq}}=\begin{cases}
A\,\tilde{x} & (k=0)\\
A\,\mathrm{arcsinh}(\tilde{x}/k) & (k\neq 0)
\end{cases}
\end{equation}
with an inverse mapping applied after sampling to recover the physical brightness-temperature space. In the main 3D comparisons, the fiducial setting uses the linear case $(A,k)=(0.1,0)$.

\subsection{Stability and Memory Controls for 3D}
The dimensional expansion from 2D to 3D required several explicit controls for training stability and memory efficiency, including distributed data parallel execution, automatic mixed precision, gradient accumulation, optional checkpointed U-Net blocks, and anisotropic downsampling choices tailored to the line-of-sight direction. The reported 3D runs were trained on 32 NVIDIA A100 GPUs with 80\,GB of device memory, and even at that memory scale the full-resolution $64\times64\times1024$ setup remained constrained to very small micro-batches. 
The training set of 25,600 lightcones is distributed evenly across 32 GPUs,
so each GPU is assigned 800 lightcones per epoch.
% In the fiducial 3D setup, each GPU processes 800 lightcones, corresponding to the full training set of 25,600 lightcones distributed across 32 GPUs. 
% We use a micro-batch size of 2, gradient accumulation of 16, and an additional amplitude-compression step \(\text{squish}(0.1,0)\), with training carried out for 120 epochs. 
\rev{The fiducial 120-epoch 3D configuration uses a micro-batch size of 2, gradient accumulation of 16, and an additional amplitude-compression step \texttt{squish(0.1, 0)}.}
These settings correspond to a cosine learning-rate schedule with $T_{\max}=3000$ optimizer steps.
\rev{The 240-epoch 3D model used in
Figures~\ref{fig:single_job_tb}--\ref{fig:single_job_scattering}
uses the same optimizer, micro-batch, and gradient-accumulation settings,
with the cosine schedule extended to $T_{\max}=6000$ optimizer steps.}
The anisotropic stride design is particularly important in 3D because it reduces memory usage while remaining better matched to the elongated lightcone geometry along the line of sight.

\bibliography{sample701}{}
\bibliographystyle{aasjournalv7}

%% This command is needed to show the entire author+affiliation list when
%% the collaboration and author truncation commands are used.  It has to
%% go at the end of the manuscript.
%\allauthors

%% Include this line if you are using the \added, \replaced, \deleted
%% commands to see a summary list of all changes at the end of the article.
%\listofchanges
\end{CJK*}
\end{document}